\documentclass{article}
\usepackage{url}
\usepackage{amsfonts}
\usepackage{natbib}
\usepackage{listings}
\newcommand{\ignore}[1]{}
\newcommand{\mod}{\,\mathrm{mod}\,}
\newcommand{\F}{\mathbb{F}}
\title{ Proofs of Zero Knowledge }
\author{Matthias Bauer \\ {\tt matthiasb@acm.org}}
\begin{document}
\ignore{ $
$Id: pozk.tex,v 1.18 2004/07/14 12:25:19 roo2b Exp $
$}
\maketitle
\begin{abstract}
\noindent
We present a protocol for verification of ``no such entry''
replies from databases. We introduce a new cryptographic 
primitive as the underlying structure, the keyed
hash tree, which is an extension of Merkle's hash tree.
We compare our scheme to Buldas et al.'s Undeniable Attesters
and Micali et al.'s Zero Knowledge Sets.
\end{abstract}


In the following, the term {\em database}  refers to a
system supplying the simplest form of databases,
a table of $(key, value)$ pairs with functions for 
keyed insertion and retrieval.

\section[Motivation]{Motivation: Untrusted Databases}
There are situations where the users of a database system
cannot trust the database. For example, the database
might be outside the users' security perimeter, or
might not follow the same security policies as the
users. The users must therefore assume that the database
could be taken over by an attacker.

The entities writing the database entries want to
be sure that the database delivers their entries
on requests unchanged. The database maintainer
wants to be able to defend itself against accusations
of misbehavior.

\section{Threat scenario}
How can an attacker or malicious database maintainer deceptively 
influence the database's replies?
The following list explains the attacks:
\begin{enumerate}
\item {\bf Changing existing entries:} The attacker can change the value
of an existing entry.\label{cha1}
\item {\bf Creating false entries:} The attacker can create $(key, value)$
pairs himself.
\item {\bf Re--labeling existing entries:} The attacker can insert 
existing values under different keys. \label{cha3}
\item {\bf Returning old entries:}  \label{cha4} If the value of a
$(key, value_1)$ pair is overwritten by a new $value_2$, the attacker
can return the old $value_1$ on request for $key$.
\item {\bf Denying existing entries:} On a $search$ request, the 
attacker can make the database deny that a $value$ exists 
for the given $key$ in the table. \label{cha5}
\end{enumerate}

This is not only of theoretical interest. 
Real--life examples are web--hosting solutions, where the
changing parts of all web pages are kept in a central relational
database and the pages are constructed from templates.
Another example is the Internet's Domain Name System (DNS),
which is basically a partially distributed lookup--table.
Attacks on DNS as listed above are described in
\cite{DNSspoof}, remedies are suggested in the 
DNS Security Extensions\cite{rfc2535} and a draft
by Bellovin \cite{Bellovin}. 

\subsection{Detecting Attacks}
The readers should be able to detect if the database is
lying about entries. The database should be able to
prove the correctness of its answers as long as the
writers are honest.
For most of the listed attacks, this can
easily be achieved if public key cryptography is employed.
It is not necessary to
assume the existence of a global Public Key Infrastructure for
this purpose. It is sufficient if the readers of the database
have access to all the public keys of the writers (presumably
a smaller set than the readers). The keys must be stored outside
the database.  With this information at hand,
and routines for signature creation ($S_k$) and verification ($V_k$),
readers and writers can make attacks \hbox{\ref{cha1} -- \ref{cha3}}
detectable
under standard cryptographic assumptions.
Here is how:
\begin{enumerate}
\item{\bf Changing existing entries:} Writers sign the values
of their entries and include the signature in the $value$ of
the table. Readers check the signatures and detect changes
by the database. $$(key, value) = (key, \mathrm{data} || S_k(\mathrm{data})) $$
\item{\bf Creating false entries:} The same applies here. Since
the database has no access to the writer's secret keys,
it cannot produce deceptive entries.
\item{\bf Re--labeling existing entries:} This calls for
an extension: the writers now include the $key$ in the
data they sign:
$$(key, value) = (key, \mathrm{data} || S_k( key || \mathrm{data} ))$$
The readers can now check if the entry belonged to the $key$
they gave in their {\em search} request.
\end{enumerate}

Points \ref{cha4} and \ref{cha5}, however, pose a harder problem.
If there is no information about existing entries anywhere
outside the database, then these attacks cannot be detected.

As for point \ref{cha4}, from the algebraic properties of
standard digital signatures it follows that a signature
on an old entry will still be valid after the entry has been
replaced. There are signature schemes
where the signer's cooperation is necessary for verification,
called {\em undeniable signatures}. The concept was presented
first by Chaum \citep{undeniable} in 1989. 
Such a scheme in our scenario would introduce much
more communication than in the standard writer--database--reader
setup. A writer who substitutes an existing entry would have
to reliably notify the writer of the previous version,
so that the previous author retracts the signature (e.g. by
sending a new ``signature outdated'' message in the verification
protocol). Readers
would need to communicate with writers who might not be
available every time someone wants to validate a reply.
We would like to avoid such a complicated and error--prone setup.

The problem with point \ref{cha5} is that
from the perspective of the database, it would mean proving
that it does {\em not know} the table entry, an impossible
feat in this scenario. Since there is no $value$ returned, there can be no
signature, unless the writers supply a special ``no entry'' value
tied to {\em every possible and unused} key.

There are ways around this, however, if we allow the writers
and readers access to another service besides the database.
The service is required to supply a single value on request;
authenticated writers must be able to overwrite that value.
We will call this service the {\em announcement service};

The writers store a signed, condensed description of the database's state.
This value must be updated after every write to the database.
The readers can use this description to validate the database's
replies.
We will call this summary of database state a {\em state credential}.
The validation protocol consists of two parts:
\begin{itemize}
\item A writer builds and publishes a new state credential from the previous one.
\item Readers check the validity of a database reply. This requires
interaction with the announcement service and the database.
\end{itemize}

\section[State Credentials]{State Credentials for Databases}
Our requirements for the state credential (to be short and
uniquely bound to the database state) suggest the application
of a cryptographic hash function. We will now describe several
possibilities to build state credentials with hash functions.

\ignore{
\subsection{Bloom Filters}
Bloom Filters \cite{Bloom} provide means to verify set--membership
{\em probablistically}. The data structure used is a simple bit--vector
of fixed length. To insert an entry into the filter, several
indices into the bit--vector are computed by application of
hash functions. The bits corresponding to the indices are
set to~$1$. To check if a given object is in the set, the
same functions are applied and the indexed bits checked. 
Should they all be~$1$, the object is considered to be 
a member. The state credential in our setting would be 
the bit--vector signed by the writer of the last entry.
Depending on the size of the bit--vector,
the output length of the hash functions and
the number of entries, false positives are possible in
Bloom Filters.  This violates our requirement that the database must
always be able to prove the correctness of its replies.
}

\subsection{Simple Hash}
This is the simplest form of a state credential. The writer
queries the database for all entries and inserts the 
{\em key} of his/her new entry. The writer then sorts all
keys in a pre--defined
order and computes the hash value over all of them. 
This value is signed and supplied through the announcement
service.
Formally: $$c = h(\sum_{i=0}^{n}k_i)$$
where $n$ is the number of entries in the database, $k_i$
is the $i$th {\em key} in the particular order
and addition denotes string concatenation
after adding a special ``end of key'' marker to the parameters.
The simple hash is not
a solution, because to check a database reply against this,
a reader would have to download the whole database as well.

\subsection{Hash Chain}
Application of hash chains make the computation of the credential 
cheaper. The writer pulls the latest credential from the
announcement service and ``adds'' the new entry's {\em key}.
Formally: $$c_0 = h() $$
$$c_i = h(c_{i-1} + k_i)$$
where $c_i$ is the credential after adding the $i$th entry
to the database and $h()$ is the hash of the empty string.
Addition is defined as above.

To validate a database reply, a reader would have to pull
all keys up to the requested key in their order of addition.
In the interesting case of a non--existing key, the
reader has to pull the whole database. So this 
is still not satisfactory.

\subsection{Hash Trees}
Merkle's \cite{hashtree} hash trees allow
fast (Log-Time) verification of digests over many entries.
 Hashes over pairs of entries 
(the keys in our scenario) are computed and the resulting
hashes are again paired and new hashes computed and so on.
The number of entries has to be a power of $2$.
Formally:
$$ H(l,j) = h(k_j) $$
$$ H(i,j) = h(H(i+1, \lceil\frac{(i+j-1)}{2}\rceil) . H(i+1, \lceil\frac{i+j+1}{2}\rceil)) $$
where $l = \log_2(\#\mathrm{entries})$, by ``$.$'' we denote the
concatenation of strings and $k_j$ is the $j$th key in
an arbitrary order ($H(1,1)$ is the root of the tree).

The state credential at the announcement service
must be updated for each new entry and a hash tree
root can only be computed if the number of entries is 
a power of two. So we must find a form of
state credential that allows instant updating but
still can use hash trees. 

The solution is to split the number of entries 
into powers of $2$ and to generate hash trees of
appropriate height for all of those powers.
For this reason, the state credential will
consist of a number of hash tree roots, 
at most $\log_2(n - 1)$ where $n$ is the number of entries,
together with the height of the trees.
When an entry is added to the non--empty database, and if the
number of entries is odd after the addition, then a
new hash tree of height zero (simply the hash of the key)
is appended to the 
state credential. If the number is even, then the two
most recent tree roots are combined  (hashed together) to form the root
of a higher tree. This is done recursively from the end.
Note that we must start at the zeroth
entry,
since even an empty database needs a state credential.

So if the bit representation is $$n - 1 = \sum_{i=0}^{\lfloor\log_2(n-1)\rfloor} b_i \cdot 2^i$$
and $t = \sum b_i$ (the Hamming--weight of $n-1$),
then the credential consists of the roots of $t$ hash
trees of height $i$ for every $b_i = 1$.

The database in turn keeps a counter of entries.
It can thus deduce  in which
of the trees in the current state credential an entry resides.
This is achieved by comparing the entry's number against
the current set of trees and their heights.

If a reader wishes to validate the databases' reply for
a certain key $k$, the database has to supply a
path in the tree leading to the leaf with $k$, as well as all pairs
of hashes along that path. The reader can now verify
that each pair's concatenation hashes to the corresponding
hash in the next pair, up to the root of the tree in the
state credential. Communication with the database for
validation is bounded at worst by $\log_2(n)\cdot 2 \cdot |h()|$~bytes,
where $|h()|$ is the byte--length of the hash's output.

The drawback of this scheme is that validation of
a negative reply (``no such entry'') still requires
downloading the whole database.

\section{Keyed Hash Trees}\label{keyedhash}
Our contribution is 
the introduction of {\em keyed hash trees}. In these, the position of
a leaf in the tree is dependent on the key corresponding
to the leaf. Our construction uses a tree of height $|h()|$,
where $h$ is a cryptographic hash function.
For every $(key, value)$ in the database, a leaf
is inserted in the tree, where the path to the leaf from
the root is defined by $h(key)$ and
the data in the leaf is $h(value)$. Leaves without
a corresponding entry are implicitly
set to the hash of the empty string $h()$.

We call a sub--tree {\em empty} if there is no path leading
through the sub--tree's root that leads to a leaf with
a non--empty value. All leaves of an empty sub--tree have
the value $h()$. The nodes in the layer above the leaves
all have value $h(h(),h())$, and so on. Note that all nodes
in a layer have the same value, which in turn is derived
from the hash of the empty word. It is easy to pre--compute
these values. It allows identification of a sub--tree as being empty 
by a single table lookup, since all empty sub--trees of
equal height have the same pre--computed root hash.

From the collision resistance for cryptographic
hash functions, it follows that $$\forall a,b: a \neq b :\, h(a . b) \neq h (b . a) .$$
The value of the tree's root thus depends on all the leaves'
values and all the paths. In contrast to the schemes above,
non--existing entries {\em do}  have leaves and paths and
can thus be validated.

Storing and working on a binary tree of height~$160$ (for $h = \mathrm{SHA1}$
for example) seems daunting at first,
since there would be $2^{160+1} - 1$ entries, only a few
orders of magnitude less then the number of hydrogen atoms
in the whole universe. But all except $n$ of the leaves are
empty, where $n$ is the number of entries in the database.
We will show in Section \ref{sparse} that it is sufficient
to store only the non--empty leaves and the branches leading to them
from the root.

Before inserting a new entry $(key, value)$ in the database, a writer
requests all pairs of hashes on the path to the new entry.
The writer then validates the hashes against the current
state credential, which is the root of the keyed hash tree.
The writer substitutes the hash in the respective leaf by the hash of the
{\em value}, and computes a new root hash. He then signs this root hash
and puts it on the announcement service after inserting the new
entry in the database.

To validate a database reply, the reader requests all
pairs of hashes on the path derived from the key.
For positive replies, the reader hashes over the returned
value, compares that with the hash in the corresponding leaf
and proceeds as with a standard hash tree.

If the reply was negative (``no such entry''), then
the path must lead to an empty leaf and  --- as in the
hash tree scheme above --- the hash values can be
checked recursively up to the root. The root
must be the same as the signed value received from
the announcement service. 

\section{Sparse Hash Tree Algorithms}
\label{sparse}
Our database should be able to respond quickly to
requests for pairs of hashes along a given path.
To do this, we need an internal representation of the
keyed hash tree corresponding to the current database table. 

We are helped by the keyed hash tree's property that all
empty sub--trees of equal height are identical
(an algorithm for creating a list of all empty sub--trees' roots
is given in figure \ref{emptys}).

Instead of storing $2^{161}-1$ nodes, we need to store only
those nodes that are part of a path to an existing
entry in the database, i.e. the hash of at least one key in the
database describes a path leading through the node.
To reduce the space further, we store only one node
for every sub--tree that contains exactly one entry,
i.e. the first node from the root down which lies
on one single path is used to describe the whole
sub--tree containing the corresponding leaf.
We call the resulting data structure a {\em sparse hash tree}.

The main problem is that searching in the
tree is done from the root down to the leaf, while the
hash values in the nodes are computed from the leaves up.

\lstset{language=C,
		tabsize=4,
		basicstyle=\tiny\ttfamily,
		frame=lines
		}
\begin{lstlisting}[label={emptys},float, 
				   caption={Pre--Computation of all empty sub--trees},
				   ]
list empty_init(void){
	i = 0;
	Empty[i] = h();
	for (;i<159;i++) {
		Empty[i] = h(Empty[i-1] . Empty[i-1]);
	}
	return Empty;
}
\end{lstlisting}

Each stored node in our hash tree representation is a {\tt struct}
defined in listing \ref{structs}
\begin{lstlisting}[float, caption={Structures {\tt node} and {\tt entry}},
label={structs}]
struct node {
	struct node *b[2];
#define LEAF   1
#define BRANCH 2
	int flag;
	struct entry *e;
	unsigned char* hash;
};

struct entry {
	unsigned char* path;
	unsigned char* value;
};
\end{lstlisting}

The {\tt b[2]} array contains pointers to the left and right children,
or NULL if there is no respective child. 
{\tt flag} indicates whether this node represents a {\tt BRANCH}
or a {\tt LEAF}. {\tt hash} is the hash of the tree rooted in this
node. Only in leaf nodes is {\tt e} not {\tt NULL} and contains a pointer
to a {\tt struct entry} defined in listing \ref{structs}. Note
that leaf nodes do represent the leaves of the tree, but only
non--empty leaves have leaf nodes. The nodes contain data structures
that allow dynamic computation of all node--values of the hash
tree below their {\tt path}.

{\tt entry} holds the information necessary to
compute all nodes below the node pointing to it
(see algorithm {\tt leafnode} for details).
{\tt path} is a bit--string of {\tt 0}s and {\tt 1}s and
describes the path to the actual leaf from the leaf node,
i.e. the part of the path for which we don't store nodes,
but generate them as needed.
{\tt value} contains the hash of the entry at the actual
leaf.

\subsection{Computing Pairs along a Path}
The function {\tt rootpath} defined in listing~\ref{rootpath} on page \pageref{rootpath} returns a list of
pairs of hashes along a given path. The first two {\tt If}
statements handle the special cases that there is at most
one entry in the sparse tree.
\begin{lstlisting}[float,
                   caption={{\tt rootpath()}: 
					        Generating all pairs of hashes along {\tt path}},
			       label={rootpath}]
list rootpath(node root, bitstring path) {
	if(root == NULL){ 
		/* empty sub-tree */
		return nullnode(path); 
	}
	If(root->flag == LEAF){
		/* exactly one leaf in subtree */
		return leafnode(root, path);
	}
	if(root->flag == BRANCH){
		/* non-unique path */
		init (List);
		return branchnode(path, root, List);
	}
}
\end{lstlisting}
{\tt rootpath} calls the auxiliary functions {\tt nullnode},
{\tt leafnode} and {\tt branchnode}.  

{\tt nullnode} 
simply builds pairs of empty sub--tree
roots in ascending order. 

{\tt leafnode} 
calls {\tt singlepath} 
to generate a temporary list of all non--empty sub--trees below the
leaf node. It then compares the supplied path against the path
to the leaf and selects the pairs from the temporary list for
the maximum left--match of the paths. For the rest of the path,
empty sub--tree roots are appended to the selected pairs.
This list is returned.

{\tt branchnode} walks recursively down the tree as long as
the path runs through BRANCH nodes. At each step it appends the
hashes in the children of the current node to the list.
At the point where the path
changes to an undefined or LEAF node, {\tt branchnode} calls
{\tt nullnode} or {\tt leafnode} respectively, to complete
the list.

\begin{lstlisting}[float, caption={{\tt nullnode()}},label={nullnode}]
list nullnode(bitstring path){
	init (List);
	for(i=0; i < length(path); i++){
		push List, (Empty[i], Empty[i]);
	}
	return(List);
}
\end{lstlisting}
		
\begin{lstlisting}[float, caption= {{\tt singlepath()}: 
Computing pairs of hashes in a sub--tree with exactly one given {\tt  %
(path, value)}},
label={singlepath}]
list singlepath(bitstring path, char value[160]) {
	revpath = reverse (path);
	init (List);
	init (pair);
	i=1;
	bit = shift revpath;
	pair[bit] = value;
	pair[not bit] = Empty[0];
	push (List, pair);
	while(bit = shift revpath) {
		pair[bit] = h(List[0][0] . List[0][1]);
		pair[not bit] = Empty[i++];
		push (List, pair);
	}
	return List;
}
\end{lstlisting}
	
\begin{lstlisting}[float,caption={{\tt leafnode()}: 
Generating pairs of hashes along a {\tt path} below a leaf {%
\tt node}},label={leafnode}]
list leafnode (struct node *node, bitstring path) {
	init (TmpList);
	init (List);
	init (pair);
	i=0;
	/* create list of non-empty nodes' hashes */
	TmpList = singlepath(node->e->path, node->e->value);

	/* compare given path against path of leaf */
	while(path[i] == node->e->path[i]) {
		append (List, TempList[i]);
		i++;
	}
	height = length(path) - i;

	/* if the two diverge, return empty nodes' hashes */
	while(height > 0) {
			pair = (Empty[height], Empty[height]);
			append (List, pair);
			height--;
	}
	return List;
\end{lstlisting}

\begin{lstlisting}[float,caption= {{\tt branchnode()}: 
Generating pairs of hashes along a {\tt path} below a branch {\tt node}},label= {branchnode}]
list branchnode (struct node *n, bitstring path) {
	init (pair);
	/* get hashes of children */
	for(b = 0, 1){
		if (n->b[b] != NULL){
			pair[b] = n->b[b]->hash;
		} else {
			pair[b] = Empty[length(path)];
		}
	}
	append (List, pair);
	bit = shift path;
	
	if (n->b[bit] != NULL && n->b[bit]->flag == BRANCH){
		/* walk down */
		append (List, branchnode(path, n->b[bit], List));
	} else {
		if(n->n[bit] != NULL && n->b[bit]->flag == LEAF){
			/* we hit a leaf */
			append (List, leafnode(path, n->b[bit]));
		} else {
			/* walked into an empty sub-tree */
			append (List, nullnode(path));
		}
	}
	return List;
}	
\end{lstlisting}

\subsection{Inserting an Entry}
{\tt insert} (algorithm \ref{insert} on page \pageref{insert}) 
recursively walks down the tree along the given path,
putting the traversed nodes on the stack. If the path runs into
an empty subtree, a LEAF node with the given (path, value) is
created. If the path runs through a LEAF node, the {\tt entry} in
the node is moved one step along its path down the tree, the node
is converted to a BRANCH, and {\tt insert} is called again
on it. On the way up to the root, all the hashes on the path 
are adjusted.

\begin{lstlisting}[ label={insert},float=h,caption= {{\tt insert()}: Insertion of a new entry}]
struct node *insert(bitstring path, char value[160], struct node *node) {
	if (node == NULL ){
		/* virgin sub-tree: create new leaf */
		init (List);
		init (newnode);
		init (entry);
		newnode->flag = LEAF;
		List = leafnode(path, value);
		entry->path = path;
		entry->value = value;
		newnode->e = entry;
		newnode->hash = h(List[0][0] . List[0][1]);
		free(List);
		return(newnode);
	}
	if(node->flag == BRANCH){
		init (newnode);
		bit = shift path;
		/* walk down */
		newnode = insert(path, value, node->b[bit]);
		/* update the hash value on the way back */
		init (pair);
		pair[bit] = newnode->hash;
		if (node->b[not bit] == NULL) {
				pair[not bit] = Empty[length(path)];
		} else {
			pair[not bit] = node->b[not bit]->hash;
		}
		node->hash = h(pair[0] . pair[1]);
		return(node);
	}
	if(node->flag == LEAF){
		/* change LEAF to BRANCH */
		init (tmppath);
		init (tmpval);
		tmppath = node->e->path;
		tmpval = node->e->value;
		node->flag = BRANCH;
		/* move leaf's content down along its own path */
		free node->entry;
		node->entry = NULL;
		node = insert(tmppath, tmpval, node);
		/* insert new (path, value) in the now BRANCH */
		bit = shift path;
		init (newnode);
		newnode = insert(path, value, node->b[bit]);
		/* update hash on the way back */
		node->b[bit] = newnode;
		init (pair);
		pair[bit] = newnode->hash;
		if (node->b[not bit] == NULL) {
		    pair[not bit] = Empty[length(path)];
		} else {
			pair[not bit] = node->b[not bit]->hash;
		}
		node->hash = h(pair[0] . pair[1]);
		return(node);
	}
}	
\end{lstlisting}

\subsection{Deleting an Entry}
{\tt delete} (algorithm \ref{delete} on page \pageref{delete}) recursively  walks down the tree until it reaches
the LEAF node with the entry to delete. On the way up, it checks
if the current node has only one child, which is a LEAF. If so, the
{\tt entry} in the leaf is attached to the parent, and the child deleted.
This is to handle situations  where
two leaves hang at the end of a stalk. If one of the leaves is
removed, the stalk would be a series of roots of subtrees with
exactly one entry. To keep storage minimal, this should be
avoided. {\tt delete} attaches the lonely leaf at the uppermost branch 
of the stalk.

\begin{lstlisting}[float=h, caption={{\tt delete()}: Deleting the entry at the end of a given path}, label={delete}]
struct node *delete(bitstring path, struct node *node){
	if(node->flag == LEAF) {
		/* remove LEAF, stop recursion */
		free (node->entry);
		free (node);
		return NULL;
	}
	/* we're on the way to the leaf still */
	init (pair);
	/* remember the bit */
	bit = shift path;

	/* walk down */
	node->b[bit] = delete(path, node->b[bit]);

	/* reconstruct the path */
	path = prepend bit, path;

	/* how many non-children ?*/
	for (i = 0, 1) {
		if(node->b[i] == NULL) {
			emp = i;
			numemp++;
		}
	}
	if(numemp == 2){
		/* node was a stalk with a single leaf, delete */
		free (node);
		return NULL;
	}
	if(numemp == 1){
		if(node->b[not emp]->flag == LEAF){
			/* one child, a leaf, move it up */
			node->e = node->b[not emp]->e;
			node->e->path = node->e->path;
			/* correct the path */
			pop node->e->path; 
			node->flag = LEAF;
			pair[emp] = Empty[length(path)];
			pair[not emp] = node->b[not emp]->hash;
			node->hash = h(pair[0] . pair[1]);
			free (node->b[not emp]->e);
			free (node->b[not emp]);
			return node;
		} else {
			/* branch node leading to at least two leafs */
			/* update the hash */
			pair[emp] = Empty[length(path)];
			pair[not emp] = node->b[not emp]->hash;
			node->hash = h(pair[0] . pair[1]);
			return node;
		}
	}
	if(numemp == 0){
		/* branch node with at least two leafs below */
		/* update the hash */
		pair[0] = node->b[0]->hash;
		pair[1] = node->b[1]->hash;
		node->hash = h(pair[0] . pair[1]);
		return node;
	}
}
\end{lstlisting}

\subsection{Properties}
We assume that the hash function $h$ behaves
as an ideal hash function would, an assumption often used in
cryptography called the Random Oracle Model. It implies that
the output of $h$ is indistinguishable from the output of a
random function.
Under this assumption, our algorithms have the following
properties:
\begin{description}
\item {The sparse tree in memory is nearly balanced.}
If $h$'s output behaves as randomness, the paths in the tree
will be random walks starting at the root. For a large number $n$
of entries, $\frac{1}{2^{i}}$ paths will lead through each
node at the $i$th layer in the tree, and the average path
length is $\log_2(n)$.
\item {Space for the sparse tree is bounded by twice the 
number of entries in the database.} We store $n$ entries at $n$
leaves, and each pair of nodes has one parent--node. This means
that
\begin{eqnarray*}
maxnodes(n) & = & \sum_{i=0}^{\lceil \log_2(n) \rceil}{2^i} \\
			& = & 2^{\lceil \log_2(n) \rceil + 1} - 1\\
			& < & 2\cdot n \\
\end{eqnarray*}

\item {Pairs along paths to a non--existing key can be computed quickly.}
If the key is not in the table, then the path will enter an empty
subtree after $\log_2(n)$ steps, in the mean. For one million
entries, for example, this
means that after traversal of 20 nodes on average, {\tt nullnode} will
be called, and all that remains to be done is table lookups.
\item {Readers recognize non--existing keys after $\log_2(n)$ steps
from the root, in the mean.} The pre--computation shown in \ref{emptys}
consists of 160 calls to the hash function. After this, a reader
simply checks the returned pairs from the database against the 
pre--computed table {\tt Empty}.
\item {Maximum message size per validation is $2 \cdot |h()| \cdot |h()| $~bits.}
Two pairs of $|h()|$ bit hashes per step lie on a path with $|h()|$ steps.
For SHA-1 as $h$, this would mean $6400$~bytes per validation.
If stronger collision resistance is required, 
a cryptographic hash function with longer output may be chosen.
While the resistance grows exponentially with the hash size,
the messages grow only linearly.
\item {The database can be distributed over several machines.}
The {\tt insert}, {\tt delete} and {\tt rootpath} functions are independent
of the actual height of the stored tree. As long as writers
and readers know how to compute the $2^{d+1}-1$ hashes at the
root of the tree, they can use $d$ independent databases.
This would also allow $d$ concurrent writes, by locking at
the branches.
\end{description}

\section[Related Work]{Related Work}
There are other schemes with different objectives related
to the one just presented. We will discuss the differences
to our proposal. 

\subsection{Undeniable Attesters}
\label{buldas}
Ahto Buldas et al. presented schemes for
accountable certificate management in 2000 (\cite{buldasold} and
\cite{buldas}). In their scenario,
the database is a part of a certificate authority (CA), i.e.~the
list of valid certificates. Their goal is to make sure that
there is never an ambiguity about the state of a certificate.
On request for a key $k$, the database sends an {\em attester} 
$p = P(k,S)$, which
is a cryptographic statement about the presence or absence
of the key  $k$ in the table $S$. 
The database also returns a {\em digest} $d = D(S)$ which is a 
summary of the database's table $S$.
Any party can then call a verification algorithm
$V(k,d,p)$ which returns ``Accept'' if the statement $p$
about $k$ was correct for table summary $d$ or else ``Reject''.
Buldas et al. call
an attester {\em undeniable} if a CA can produce
two contradicting attesters for the same key with only 
negligible probability. Formally, this is defined as follows:

Let $\cal{EA}$ be the class of probabilistic algorithms of polynomial
runtime. Let $k$ be the security parameter (in our context
the bitlength of the hash function's output). The attester $\cal{A}$
is given by the tuple of algorithms $\cal{A} = (G, P, D, V)$%
\footnote{$G$ serves only to select a hash function for a 
given security parameter $k$.}. Its resilience against an
attacker $A$ of class $\cal{EA}$ is defined as
$$ \mathrm{UN}_{\cal{A},k}(A) = \mathrm{Pr}[(x, d, p, \bar p ) 
	\leftarrow A(1^k): V(x,d,p) = \mathrm{Accept} \wedge 
					   V(x,d,\bar p ) = \mathrm{Reject}]\,.$$
If $k$ can be chosen as to minimize $\mathrm{UN}_{\cal{A},k}$ below
any given $\epsilon$, then $\cal{A}$ is called undeniable.

The paper 
examines different schemes for attesters and concludes 
with an efficient, undeniable attester based on search trees.

\subsubsection{Search Trees}
A search tree (see for example Knuth \cite{Knuthsearch}) is a
binary tree with the additional property that
there is a comparison relation $<$ and each
node contains a value $v$ for which the following holds:
$$ v_l < v < v_r ,$$
where $v_l$ and $v_r$ are the values in the  left
and right child of the node, respectively. If one or both children
do not exist, the empty string is used instead. Buldas et al. add
another field to each node, which holds the hash
$h_v = h(h_{v_l} . v . h_{v_r})$. 
The  digest  $d$ --- the root of the search tree ---
is signed by the CA and 
published through an untrusted Publication Authority (PA).
The digest corresponds to our state credential and 
the PA to our announcement service. 

A proof $p$ output by $P$  for a key $k$ is a list $p = (V_0, \dots, V_m)$ of values $V_i =(h_L,k_i,h_R)$,
where $k_i$ is the value of a node and $h_L, h_R$ are the hashes
stored in the left andx. right child of the node, repectively, or the empty
string if there is no such node. The values are selected such 
that $k_0 = k$ if $k$ is in the database. If $k$ is not in
the database, then the tree contains two keys $j$  and $l$
such that $l$ is the smallest value larger than $k$
and $j$ the largest value smaller than $k$. By construction
of the tree, there is a path from the node of $j$ to
the root leading through $l$ or vice versa. The 
key of $j$ and $l$ which is lower in the tree is used
as $k_0$ in that case.
If $k$ is larger or smaller
than all keys in the search tree, then 
$l$ or  $j$ is set to the largest or smallest key
in the tree, respectively.

$p$ together with the published
digest $d$ allows to prove the absence of a key. 
The verifier can check the hashes from $k_0$ to the
root and verify the strict order of child nodes.
If a key $k$ is not in the tree, then $p$ will contain
$j, l$ such that 
$j < k < l \wedge \neg \exists k_i \in p: j<k_i<k \vee
k<k_i<l$. By the strict
order and lack of right children of $j$ and left
children of $l$, $k$ cannot be a node's value in the 
tree.

The verifier $V(x,d,p)$ returns ``Error'' if any of
the following fails:
\begin{itemize}
\item Checking the order of keys in $p$.
\item Computing and comparing the hashes along the path given in
$p$ up to the root.
\item Comparing $d$ against the root hash in $p$.
\end{itemize}
If $x$ is the first key in $p$, $V$ returns ``Accept'',
else ``Reject''.

\subsubsection{Differences to our Proposal}
The attester states absence or presence in a list,
exclusively. A state credential includes the {\em value}
of the entry under the {\em key}, so that changes of an
entry can be expressed.

Buldas et al. assume that the database is
reigned by a single entity which is trustworthy
at the moment when an attester is issued, but may
turn untrusted or unavailable later. That the database
signs the digest itself establishes a different scenario
than in our setting.

To show that
the database cheated, a user has to find another
attester contradicting the attester he/she received.
In our 
setting, a reader can prove that the database cheated
immediately and without contacting other readers.

\subsubsection{Reduction of State Credentials to Attesters}
We can build undeniable attesters for keys from signed state credentials.
Since values of entries are irrelevant for attesters, we will 
substitute the $(key, value)$ pairs in our algorithms by
$(key, key)$ pairs. Since the empty word might be an entry's
key in the generality of the proof,
we will use the reserved key $\tau$ instead in that case.
The digest $d$ is the state credential, but is supplied by
the database instead of the announcement service.
Our $(P, D, V)$ are defined as follows:
\begin{description}
\item $P(x,S)$ is the list of pairs of hashes along the path
given by $h(x)$ in the keyed hash tree generated from all
entries in $S$.
\item $D(S)$ is the root of said keyed hash tree.
\item $V(x, d, p)$ outputs ``Error'' if any of the following
fails:
\begin{itemize}
\item Comparing the hash at the leaf corresponding to $x$ against
$h(x)$ or $h()$.
\item Hashing the pairs of hashes up along the path $h(x)$. 
\item Comparing the resulting root hash against $d$.
\end{itemize}

\noindent
If the leaf value at the end of path $h(x)$ is the empty hash, then
$V$ returns ``Reject''. If the leaf value is $h(x)$, then
``Accept''.
\end{description}

\subsubsection{Proof of Undeniability}
Assume that some $A\in\cal{EA}$ outputs $(x,d,p,\bar p )$
with probability $\epsilon$,
such that $V(x,d,p) = \mathrm{Accept}$ and 
$V(x,d,\bar p ) = \mathrm{Reject}$. 
Since $V$ did not
return ``Error'', all the pairs of hashes in $p$ {\em and}
in $\bar p $ resolved up to $d$. This means that
$A$ produced a collision in the hash $h$ with
probability $\epsilon$, and the
values $x_1, x_2, y_1, y_2 : (x_1 . x_2) \neq (y_1 . y_2): h(x_1 . x_2) = h(y_1 . y_2)$ 
are members of $p$ and $\bar p $.
Thus we have reduced the security of our attester to the
collision resistance of the hash $h$.

\subsubsection{Considerations about the underlying Data Structures}
Bulda's et al. use the search keys as they are,
because the proofs of non--membership in the table
rely on the greater--than relation between the keys.
This has the disadvantage that the tree becomes 
unbalanced if the inserted entries are not uniformly
distributed or if the first entry (the subsequent root)
is not near the median of the set of entries. 
Unbalanced search trees cause longer 
searches, as more than $\log_2(n)$ nodes need to be
traversed for some keys.
The paper does not explain whether there is any
re--balancing (see for example \cite{Knuthsearch})
done on the search tree. Re--balancing this particular
tree type would change all the hash--entries on the
path from the previous root to the new one. Our sparse hash
tree in memory is always balanced for large $n$.

Using the keys as they are makes the attester's size
unpredictable. With state credentials, the size is fixed (and quite
small).

\subsection{Zero-Knowledge Sets}
Unknown to this author,  S. Micali, M. Rabin and J. Kilian presented a
more general scheme, called Zero--Knowledge Sets \citep{zks} in
2003. 
Their goal was to prove set--membership non--interactively and that
without leaking any other information about the set.
The data structure underlying their
scheme is almost identical to the keyed hash trees defined above.

Micali et al. use a commitment scheme to bind the database to
its previous statements about its contents. 

\subsubsection{Pedersen's Commitment Scheme}
In commitment schemes, a prover $P$ {\em commits} to a 
message $m$ by making public a commitment string $c$ and
computing a proof $r$. Before $P$ publishes $m$ or $r$,
nothing should be inducable about the message $m$ from $c$.
After $P$ sends message $m$, a verifier $V$ can check
whether this was the message committed to by $c$. For
this, $P$ supplies $V$ with $r$. A commitment scheme
is secure if $P$ can produce contradicting proofs $r, r'$
only with neglegible probability.

Pedersen's scheme requires four public parameters, i.e., two
primes $p,q$ such that $q|p-1$, and two generators $g, h$,
where $g$ and $h$ generate $G \subset \F_p^{\star}$, the subgroup
of order $q$ in $\F_p^{\star}$.
$P$ commits to $m$ by publishing
$c = g^m h^r \mod p$. $r$ is the corresponding proof, a randomly
chosen value $\le q$.
To verify the commitment, $V$ re--computes $c$ for herself and
compares. 

To produce two proofs $r, r'$ for differing messages $m, m'$,
$P$ would have to find $r, r'$ such that $g^m h^r = g^{m'} h^{r'} \mod p$.
Assume she succeeded. She then could compute $g^{m - m'} = h^{r' - r} \mod p$
and from this $(g^{m-m'})^{r'-r} = (h^{r'-r})^{r-r'} = h \mod p$  and thus
get $log_g(h) \mod p$, breaking the discrete logarithm mod~$p$.
It follows that the security of the Pedersen commitment is reducible to
the discrete log problem. It also follows that if $log_g(h)$ is
known to the prover, she can fake arbitary commitments.

Micali et al. use this scheme to construct a hash function
$H: \{0,1\}^{\star} \rightarrow \{0,1\}^k$ where $k$ is the
bitlength of prime $p$. They use this hash function to process
a data--structure very much like our keyed hash tree from
section \ref{keyedhash}.

Their prover builds the tree by first inserting
all leaves derived from the database's entries, such that the value in 
the leaf is $H(D(x))$, where $D(x)$ is the value stored in
the database under key $x$, and the leaf's position in the tree
is determined by the path described by $H(x)$. The prover
adds all nodes on those paths to the root, their values
remain undefined until later. She adds those nodes whose
parents are now in the tree, but does not repeat this recursively.
These nodes correspond to the empty subtree's roots in our
construction.

In the next step, the prover generates a random exponent $e_v$ for
every node $v$ in the structure, and stores $h_v = h^{e_v}$ in it if $v$
is a non--empty leaf or the ancestor of one, and $h_v = g^{e_v}$ for
the empty leaves. Thus the prover knows $log_g{h_v}$ for these nodes,
a property exploited later.

The prover now computes commitments for the leaves
by computing $c_v = g^{m_v} h_v^{r_v}$, where $m_v = H(D(x))$
or $0$,
$h_v$ the value stored in the last step, and $r_v$ a random
value chosen per leaf.

In the last step, the internal nodes of
the tree get their commitment values generated. The process runs from
the leaves upwards as in Merkle's scheme. Each internal node $u$
is associated a value $m_u = H(c_{u0}, h_{u0}, c_{u1}, h_{u1})$,
where $u0, u1$ are the left and right children of node $u$.
From this and the value $h_u$ stored alread in $u$, $c_u$ is
computed as $g^{m_u} h^{r_u}$ for a randomly chosen $r_u \in \F_p^{\star}$.
The commitment value for the whole database is the commitment of the
tree's root $c_D$. This value is published before any queries are
answered.

To prove that the value stored under key~$x$ is $y$ in the 
database, the prover provides the tuples $(m_v, e_v, h_v, r_v)$
for every node on the path through the tree given by $H(x)$,
together with the values $c_{v0}, h_{v0}, c_{v1}, h_{v1}$ for $v$'s siblings.

To check this proof, verifier $V$ 
\begin{enumerate}
\item compares the leaf's $m_v$ against $H(y)$.
\item checks recursively that $m_v = H(c_{v0}, h_{v0}, c_{v1}, h_{v1})$
\item checks for every $v$ on the path that $h_v = h^{e_v}$ and
$c_v = g^{m_v} h_v^{r_v}$.
\end{enumerate}

For a proof of the non--existance of key $x$ in the database,
the prover provides all the nodes' values along the path given
by $H(x)$ as long as these nodes are in the generated tree.
The last of these nodes, $u$, will have $m_u = 0$. If $P$ would
supply this to $V$, then the verifier would learn
that there are no non--empty nodes in this sub--tree.
To disguise this fact, $P$ generates a branch of non--empty nodes
down to a virtual, empty leaf (the computation goes
bottom--up however). Node $u$ becomes the branch's root for
this proof, the value $m_u = 0$ is substituted by the value
in the uppermost node of the freshly generated branch.

The prover can do this convincingly
because she can re--commit to any $(m_u, r_u)$ in node $u$.
This is caused by the special construction of $h_u$ for
empty leaves, where $P$ knows the value of $log_g(h_u)$
by design. This allows to ``glue'' the generated branch
of nodes, complete with their commitments, to node $u$.

The verification runs as above, except that $V$ checks
for $m_{H(x)} = 0$.

\subsubsection{Differences to our proposal}
In our proposal, the writers are different entities from the
database, but they compute the committment string. 
They are continuously updating and changing entries.
In Micali et al.'s terminology, our writers {\em commit} to
the history of their changes to the database, while the database
later {\em proves} that it did reply in accordance to it state
after the latest update. Zero--knowledge was no requirement in
our design.
Since the motivations and requirements are different, the
mechanisms behind
Micali et al.'s scheme and ours differ in several points:

\begin{description}
\item {\bf Multiple, mutually trusting authors} in our design,
multiple database authors (writers) were a prerequisit. It is
not obvious, how zero knowledge sets could be adapted for
multiple writers.
Two writers would have to communicate at least the secret $e_v, r_v$ for
each node to be able to insert entries. This would require a secure
authenticated channel between all writers.
\item {\bf Zero knowledge} the zero--knowledge requirement is violated in our
scheme, because a proof about any key will leak information
about the existence or non--existence of other keys. 
\item {\bf Database modification}  one of our requirements was that the database can be 
modified by the writers who can re--compute the credential/attester/commitment
for the database. 
The zero-knowledge property of Micali et al.'s scheme is
lost if the database is allowed to change/add/remove entries.
This is because the commitment values change along the path
to a modified node and thus give away information about the
minimum number of modified entries. This makes their construction
extremely static and not suitable for our application.
Micali et al. mention this in their open problems section.
\item {\bf Speed}
because Micali et al.'s scheme must satisfy additional constraints,
it is much slower (five modular exponentations per node versus
one application of an optimized hash function).
\end{description}

\section{Summary}
We have presented an efficient mechanism that allows readers to verify the
replies of a database with the help of the writers. 
To achieve this, we introduced a new cryptographic primitive,
an extension of Merkle's hash trees.
A verification protocol between reader and database
validates a reply with a low amount of data traffic.
This extends even to those replies
where the database denies the existence of the requested entry.
The mechanism thus allows a database to prove that it does
not know about such an entry. 
We examined the relationship to work in the
area of certificate management, and showed how our
proposal can be applied there as well. In
comparison to the more general Zero--knowledge sets
by Micali et al., our scheme has the advantage that
it is faster and allows subsequent modifications by multiple writers.

\subsubsection*{Acknowledgments}
The author thanks Jonathan Katz for pointing him to Micali, Rabin and Kilian's
paper, and Silvio Micali for helpful suggestions and comments.

\end{document}